\documentclass[11pt,a4paper]{article}
\usepackage{jheppub,epsfig,multicol,bbm,latexsym,graphicx,subfigure}

\title{Sun Heated MeV-scale Dark Matter and the XENON1T Electron Recoil Excess}

\author{Yifan Chen$^{a}$, Ming-Yang Cui$^{b}$,
Jing Shu$^{a,c,d,e,f,g}$, Xiao Xue$^{a,c}$,
Guan-Wen Yuan$^{b,h}$, and Qiang Yuan$^{b,f,h}$}

\affiliation{
$^a$CAS Key Laboratory of Theoretical Physics, Insitute of Theoretical Physics, Chinese Academy of Sciences, Beijing 100190, China\\
$^b$Key Laboratory of Dark Matter and Space Astronomy, Purple Mountain Observatory, Chinese Academy of Sciences, Nanjing 210023, China\\
$^c$School of Physical Sciences, University of Chinese Academy of Sciences, Beijing 100049, China\\
$^d$CAS Center for Excellence in Particle Physics, Beijing 100049, China\\
$^e$School of Fundamental Physics and Mathematical Sciences, Hangzhou Institute for Advanced Study, University of Chinese Academy of Sciences, Hangzhou 310024, China\\
$^f$Center for High Energy Physics, Peking University, Beijing 100871, China\\
$^g$International Center for Theoretical Physics Asia-Pacific, Beijing/Hanzhou, China\\
$^h$School of Astronomy and Space Science, University of Science and Technology of China, Hefei 230026, China}
\emailAdd{yifan.chen@itp.ac.cn}
\emailAdd{mycui@pmo.ac.cn}
\emailAdd{jshu@itp.ac.cn}
\emailAdd{xuexiao@itp.ac.cn}
\emailAdd{yuangw@pmo.ac.cn}
\emailAdd{yuanq@pmo.ac.cn}

\abstract{The XENON1T collaboration reported an excess of the low-energy 
electron recoil events between 1 and 7 keV. We explore the possibility to 
explain such an anomaly by the MeV-scale dark matter (DM) heated by the 
interior of the Sun due to the same DM-electron interaction as in the 
detector. The kinetic energies of heated DM particles can reach a few 
keV, and can potentially account for the excess signals detected by 
XENON1T. We study different form factors of the DM-electron interactions, 
$F(q)\propto q^i$ with $i=0,1,2$ and $q$ being the momentum exchange, 
and find that for all these cases the inclusion of the Sun-heated DM 
component improves the fit to the XENON1T data. The inferred DM-electron 
scattering cross section (at $q=\alpha m_e$ where $\alpha$ is the fine 
structure constant and $m_e$ is electron mass) is from 
$\sim 10^{-38}$~cm$^2$ (for $i=0$) to $\sim 10^{-42}$~cm$^2$ (for $i=2$). 
We also derive constraints on the DM-electron cross sections for different 
form factors, which are stronger than previous results with similar 
assumptions. We emphasize that the Sun-heated DM scenario relies on 
the minimum assumption on DM models, which serves as a general 
explanation of the XENON1T anomaly via DM-electron interaction. 
The spectrum of the Sun-heated DM is typically soft comparing to 
other boosted DM, so the small recoil events are expected to be 
abundant in this scenario. More sensitive direct detection experiments 
with lower thresholds can possibly distinguish this scenario with other 
boosted DM models or solar axion models.}


\arxivnumber{2006.12447}

\begin{document}
\maketitle
\flushbottom

\section{Introduction}
The direct detection of dark matter (DM) has reached unprecedented 
sensitivities. Nevertheless, no convincing signals have been detected 
yet (see e.g., \cite{Liu:2017drf,Schumann:2019eaa}). 
Very recently, the XENON1T collaboration reported a potential excess of
electron recoils in the range of $1-7$ keV above the known backgrounds
\cite{Aprile:2020tmw}. The total number of events in such a recoil
energy window is 285, while the expected background number is $232\pm15$,
which suggests a significance of $3.5\sigma$. Although the unknown
backgrounds from tritium decay cannot be reliably ruled out, the
estimated tritium concentration is much lower than that required to 
fit the data \cite{Aprile:2020tmw}. The subsequent search for similar
signals with the PandaX-II data gave constraints which were consistent
with the XENON1T result, although no significant signal was detected
in Pandax-II \cite{Zhou:2020bvf}.
It has been postulated that the hypothetical effects from e.g., solar 
axions \cite{Redondo:2013wwa,Moriyama:1995bz} or the neutrino magnetic moment 
\cite{Bell:2005kz,Bell:2006wi} can account for the XENON1T data. 
However, the required model parameters are found to be in conflict with 
other constraints, particularly the astrophysical observations 
\cite{Viaux:2013lha,Ayala:2014pea,Giannotti:2017hny,Corsico:2014mpa,
Diaz:2019kim} (see however, \cite{Gao:2020wer}). Alternatively, several 
attempts \cite{Smirnov:2020zwf,Takahashi:2020bpq,Kannike:2020agf,
Fornal:2020npv,Alonso-Alvarez:2020cdv,Boehm:2020ltd,Su:2020zny,Du:2020ybt} 
have been proposed to explain the XENON1T data.

While the traditional weakly interacting massive particles in the
Galactic halo are difficult to account for the XENON1T excess due to
the very low energy deposits when scattering with electrons, one 
class of models with DM being boosted to relatively high velocities 
($\sim0.1c$) can potentially work \cite{Kannike:2020agf,Fornal:2020npv}.
In Ref.~\cite{Kannike:2020agf} a fast DM component is simply assumed,
and the possible mechanisms to produce such fast DM have been discussed,
including e.g., a fast-moving subhalo, semi-annihilating DM, or nearby 
axion stars. A realization of the boosted DM scenario has been given in 
Ref.~\cite{Fornal:2020npv}, where a faster DM component from the 
semi-annihilation of DM in the Galactic center has been proposed. 
The Sun could also be a site to accumulate enough DM in its interior
via the DM-nucleon or DM-electron scattering. However, in this case, 
the required cross section is too high that the Sun would be opaque 
for the DM to escape \cite{Fornal:2020npv}.

The high-temperature plasma inside the Sun can be a natural source to heat 
up light DM particles \cite{An:2017ojc}. The temperature of the interior of 
the Sun is about $1.5\times10^7$~K. As long as the scattering between DM and 
the electrons is moderately efficient (for example, the scattering cross 
section $\sim$~pb), the DM can be heated up to energies of $\sim$keV and 
can potentially account for
the XENON1T excess. Comparing with other boosted DM models (e.g., those 
discussed in \cite{Kannike:2020agf,Fornal:2020npv}), this scenario is quite 
clear and simple: the DM-electron scattering as seen in the detector occurs 
inevitably in the Sun (or any other places with material). This model gives a 
natural boost of DM, without additional assumptions (e.g., the high-speed 
DM subhalos \cite{Kannike:2020agf}, and the semi-annihilation/multi-component 
DM \cite{Fornal:2020npv}). In particular, the heated DM from the Sun could be 
a unique signal for future tests with directional direct detection experiments.
 
\section{Dark matter heated by the Sun}

The heated DM flux observed on the Earth can be estimated as \cite{An:2017ojc}
\begin{equation}
\Phi_{\rm heat}\sim\frac{\Phi_{\rm halo}S_g}{4\pi d^2}\times\left\{
\begin{array}{ll}
\frac{4\pi R_{\rm core}^3}{3\lambda}, & R_{\rm core}\ll \lambda\\
\pi R_{\rm scatt}^2, & R_{\rm core}\gg \lambda
\end{array}
\right.,\label{phiheat}
\end{equation}
where $\Phi_{\rm halo}$ is the DM flux in the local Milky Way halo,
$R_{\rm core}\sim 0.2R_{\odot}$ is the core radius of the Sun,
$d\equiv1.5\times10^{13}$~cm is the Sun-Earth distance,
$S_g$ describes the gravitational focusing effect which enhances
the scatterings, $R_{\rm scatt}$ is the characteristic scattering 
radius at which the DM-electron scattering once on average, $\lambda$ 
is the mean free path of a DM particle inside the core of the Sun.
For weak scattering limit, the mean free path is large, and the 
scattering probability is proportional to the volume of the Sun's core, 
which is $4\pi R_{\rm core}^3/3$. On the other hand, if the scattering 
is frequent, the scattering probability is proportional to the 
characteristic scattering area $\pi R_{\rm scatt}^2$.
Note that if interaction is strong enough, $R_{\rm scatt}$ reaches
a maximum of $R_{\odot}$, and the heated DM fluxes become weakly
dependent of the scattering cross section. The factor $S_g$ is 
estimated to be ${\mathcal O}(10)$ according to the ratio of the 
escape velocity of the Sun and the halo DM velocity \cite{An:2017ojc}.

More accurately, the differential scattering cross section between
a DM particle and an electron can be written as
\begin{equation}
\frac{d\sigma}{d\cos{\theta}\ d\phi} = \frac{\bar{\sigma}_\textrm{e}}{4\pi} 
|F(q)|^2,
\end{equation}
where $\theta$ and $\phi$ represent the spherical coordinate angles 
of the final state in the center-of-mass frame 
of each scattering. $\bar{\sigma}_{e} \equiv \frac{\mu^{2}
\left|\overline{\left|\mathcal{M}_{\text {free }}\left(\alpha m_{e}\right)
\right|^{2}}\right|}{16 \pi m_{\rm dm}^{2} m_{\rm e}^{2}}$ 
is the reference electron-DM cross section evaluated at 
$q = \alpha m_\textrm{e}$, in which $m_{\rm e}$ and $m_{\rm dm}$ 
are the masses of electron and DM and $\mu$ 
is the reduced mass of the two. $F(q)$ is the DM form 
factor depending on the momentum transfer $q$. In the center-of-mass frame, 
$F(q)$ represents a weight of angle between the initial and final states. 
In this work, we consider three examples of form factors: 
$F = 1$, $F = \frac{q}{\alpha m_\textrm{e}}$ and 
$F = (\frac{q}{\alpha m_\textrm{e}})^2$. 
These operators all originate from a heavy mediator whose mass 
is much heavier than the typical momentum transfer. The latter 
two $q-$dependent form factors can be derived, for example, from 
dimension six operators with scalar-pseudoscalar interaction 
$\bar{\chi} \gamma_{5} \chi \bar{e} e$ and pseudoscalar-pseudoscalar 
interaction $\bar{\chi} \gamma_{5} \chi \bar{e} \gamma_5 e$, 
respectively \cite{Bloch:2020uzh}. 

To properly handle the multiple scatterings, Monte Carlo simulations 
are usually required to calculate the energy distribution and fluxes 
of DM reflected by the Sun. Following Ref.~\cite{An:2017ojc},
we do such simulations taking into account the updated standard solar 
model \cite{Vinyoles:2016djt}. We start the simulation through
generating halo DM particles with a truncated Maxwell distribution,
with the most probable velocity of $220$ km~s$^{-1}$ and an escape
velocity of $540$ km~s$^{-1}$. The impact parameter with respect to
the Sun is adopted to be $4R_{\odot}$. The gravitational bending of the 
DM particle outside the Sun is computed with the Newton's law. When the 
DM enters the Sun, the gravitational effect is neglected, and only the 
scattering with electrons are considered. The interior of the Sun is 
approximated to be electron-ion plasma, and the electron number density 
is calculated according to the radius-dependent composition model of the Sun 
\cite{Vinyoles:2016djt}. A mean free path of a DM particle is calculated 
as $\lambda(r)=v_{\rm dm}(r)/[n_e(r)\langle\sigma_\textrm{e} v_r\rangle]$
\cite{An:2017ojc}, which is a function of radius $r$ inside the Sun.
In the above formula, $v_{\rm dm}(r)$ is the DM velocity, and $v_r$
is the relative velocity between DM and electrons. The calculation of
the mean free path considers the fact that the thermal electrons move
with a very high speed, but do not travel far in the Sun. The interaction
probability of the DM particle is $P=1-e^{-l_{\rm step}/\lambda}$ 
after traveling a distance of one step $l_{\rm step}$. When the DM
particle leaves the Sun, a gravitational redshift of the particle energy 
with $E_{\rm final}=E_{\rm surf}-m_{\rm dm}v_{\rm esc}^2/2$ is applied,
where $E_{\rm surf}$ is the DM kinetic energy at the Sun's surface and
$E_{\rm final}$ is its final kinetic energy escaping from the Sun.
For those DM particles that do not hit the Sun, their kinetic energies keep 
unchanged as the initial values. The heated DM flux is then~\cite{An:2017ojc}
\begin{equation}
\frac{d\Phi_{\rm heat}}{dE_{\rm dm}}=\Phi_{\rm halo}\times
\frac{16\pi R_{\odot}^2F(E_{\rm dm})}{4\pi d^2},
\end{equation}
where $F(E_{\rm dm})$ is the normalized kinetic energy distribution of 
DM with the reflection of the Sun, and $16\pi R_{\odot}^2$ 
is the impact area adopted in the simulation.

\begin{figure}[ht] 
\centering 
\includegraphics[width=0.6\textwidth]{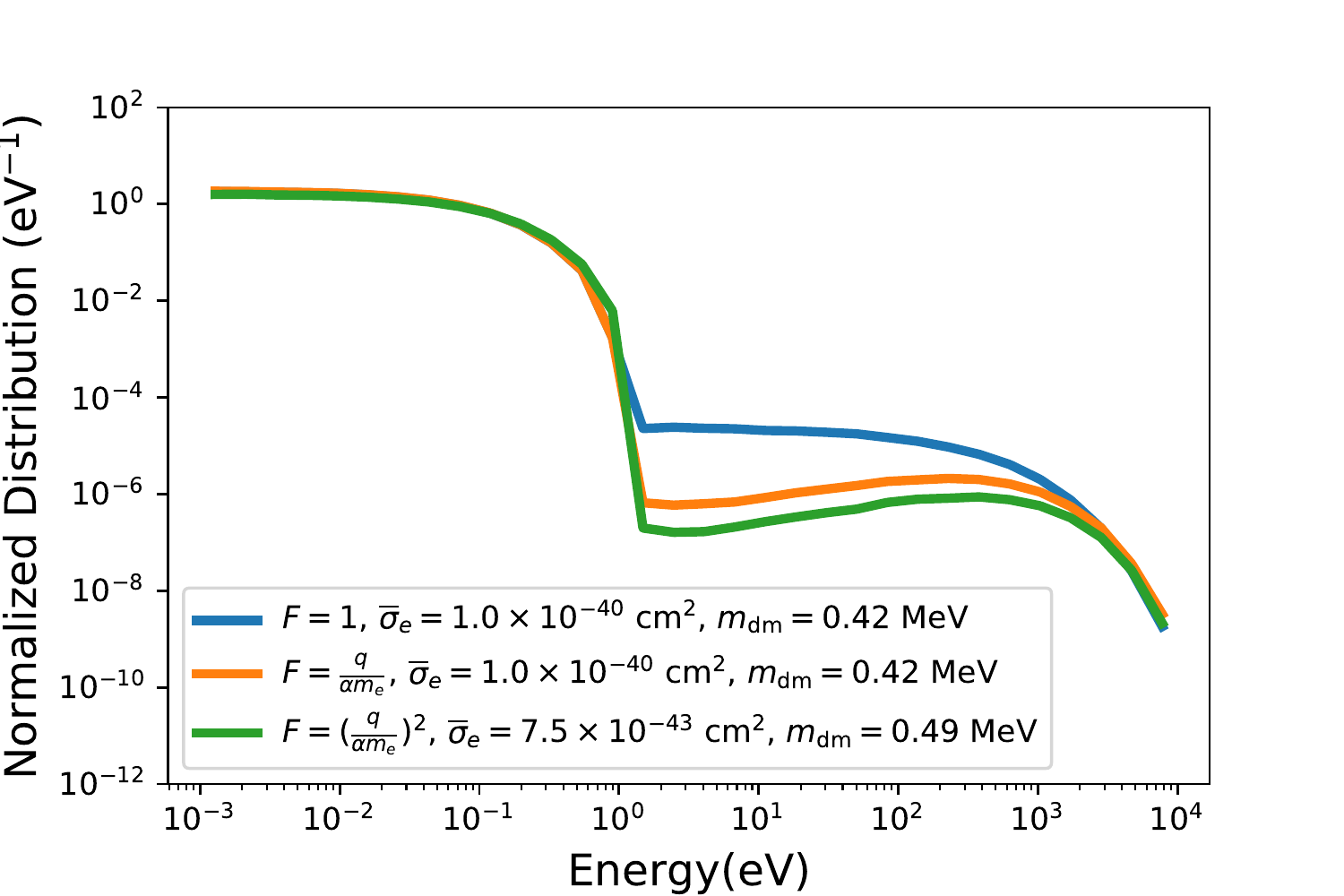}
\caption{Normalized energy distributions of the heated DM fluxes on the 
Earth for parameters given in Eq. (\ref{bf}), for $F=1$, 
$F = \frac{q}{\alpha m_\textrm{e}}$, and 
$F = (\frac{q}{\alpha m_\textrm{e}})^2$, respectively.} 
\label{bfsolarspectrum}
\end{figure}

The energy spectrum of the heated DM depends on the DM mass and the 
scattering cross section \cite{An:2017ojc}. Since most of the heated DM 
particles have kinetic energies lower than the threshold of the current
XENON1T analysis ($\sim1$~keV), we expect that only the high-energy tail 
would contribute to the XENON1T events. For a 2-body elastic scattering 
with the electron at rest, the maximum recoil energy is
\begin{equation}
E_\textrm{r}^{\textrm{max}} = \frac{4 E_{\textrm{dm}} m_\textrm{e} 
m_{\textrm{dm}}}{(m_\textrm{e} + m_{\textrm{dm}})^2} > 1~{\rm keV},
\end{equation}
where $E_{\rm dm}$ is the kinetic energy of the DM. The energy transfer 
in both the scatterings inside the Sun and in the detector is the most 
efficient if $m_{\rm dm}\sim m_{\rm e}$. 

Fig. \ref{bfsolarspectrum} shows the normalized DM energy spectrum 
after the scattering with the Sun, for the three types of DM form factors.
For each form factor, we choose the DM mass and cross section which
best-fit the XENON1T data (see below Sec. III). As expected, the higher 
power of $q$ of the form factor leads to a harder spectrum.

\section{Confronting the new XENON1T data}

To compare with the XENON1T data, we calculate the event rate 
of electron recoils in the detector as
\begin{equation}
\frac{{\rm d}N}{{\rm d}E_\textrm{r}}=N_d\times\int\frac{{\rm d}\sigma}{{\rm d}E_\textrm{r}}
(v_{\rm dm},E_\textrm{r})\,\frac{{\rm d}\Phi_{\rm heat}}{{\rm d}v_{\rm dm}}\,
{\rm d}v_{\rm dm},
\end{equation}
where $N_d\simeq4.2\times10^{27}$ ton$^{-1}$ is the number of Xe atoms 
for one ton mass of the detector, $v_{\rm dm}$ is the velocity of
the DM particle, $E_\textrm{r}$ is the electron recoil energy,  
${\rm d}\sigma/{\rm d}E_\textrm{r}$ is the differential scattering 
cross section, and ${\rm d}\Phi_{\rm heat}/{\rm d}v_{\rm dm}$ is the
spectrum of the heated DM component.  

Following Refs.~\cite{Essig:2012yx,Roberts:2016xfw,Roberts:2019chv}, the 
differential cross section for fixed DM velocity can be written as
\begin{equation}
\frac{{\rm d}\sigma}{{\rm d} E_\textrm{r}} (v_{\rm dm}, E_\textrm{r}) = \frac{\bar{\sigma}_\textrm{e} m_{\textrm{e}}}{2 \mu^2 v_{\rm dm}^2}  \int_{q_{-}}^{q_{+}} a_{0}^{2}\ q\ \mathrm{d} q\ \left|F (q)\right|^{2} K(E_\textrm{r}, q),
\end{equation}
where $a_0 = 1/(\alpha m_{\textrm{e}})$ 
is the Bohr radius, and the integration limits are $q_{\pm} = m_{\mathrm{dm}} v_{\rm dm} 
\pm \sqrt{m_{\mathrm{dm}}^{2} v_{\rm dm}^{2}-2 m_{\mathrm{dm}} E_{\textrm{r}}}$. 
The atomic excitation factor $K(E_{\rm r}, q)$, describing the probability
of obtaining a particular recoil energy for an ionized electron given
momentum transfer $q$, is taken from Ref.~\cite{Catena:2019gfa, DarkARC} 
where the initial bound states contain Roothan-Hartree-Fock (RHF) wave 
functions with the coefficients being tabulated in Ref.~\cite{Bunge} 
and the final state wave functions given in Ref.~\cite{Bethe}. 
 $K(E_{\rm r}, q)$ is related to the atomic response function 
in Ref.~\cite{Catena:2019gfa} through $K(E_\textrm{r}, q) = \sum_{n\ell} 
\frac{\left|f_{\text {ion }}^{n \ell}\left(k^{\prime}, q\right)
\right|^{2}}{2 k^{\prime 2} a_0^2} \theta (E_\textrm{r} - E_b^{n\ell})$, 
where $E_\textrm{r} = E_b^{n\ell} + k^{\prime 2} /2 m_{\textrm{e}}$, 
$E_b^{n\ell}$ is the minus binding energy of the initial state electron, 
and $k^\prime$ is the momentum of the final state ionized electron. 
We consider the contributions from all accessible atomic energy states of Xe.

\begin{figure*}[ht]
\centering 
\includegraphics[width=0.45\textwidth]{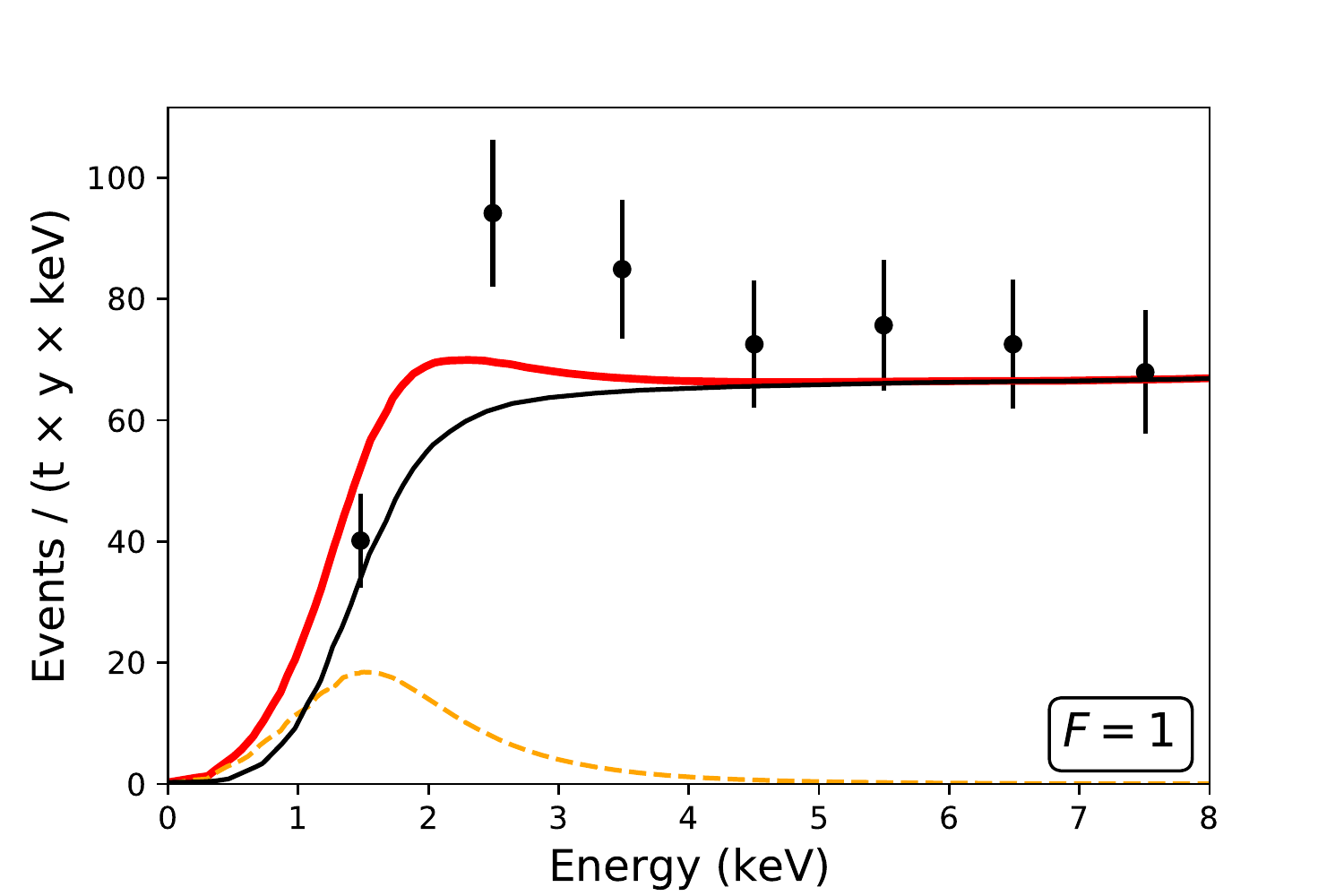} 
\includegraphics[width=0.45\textwidth]{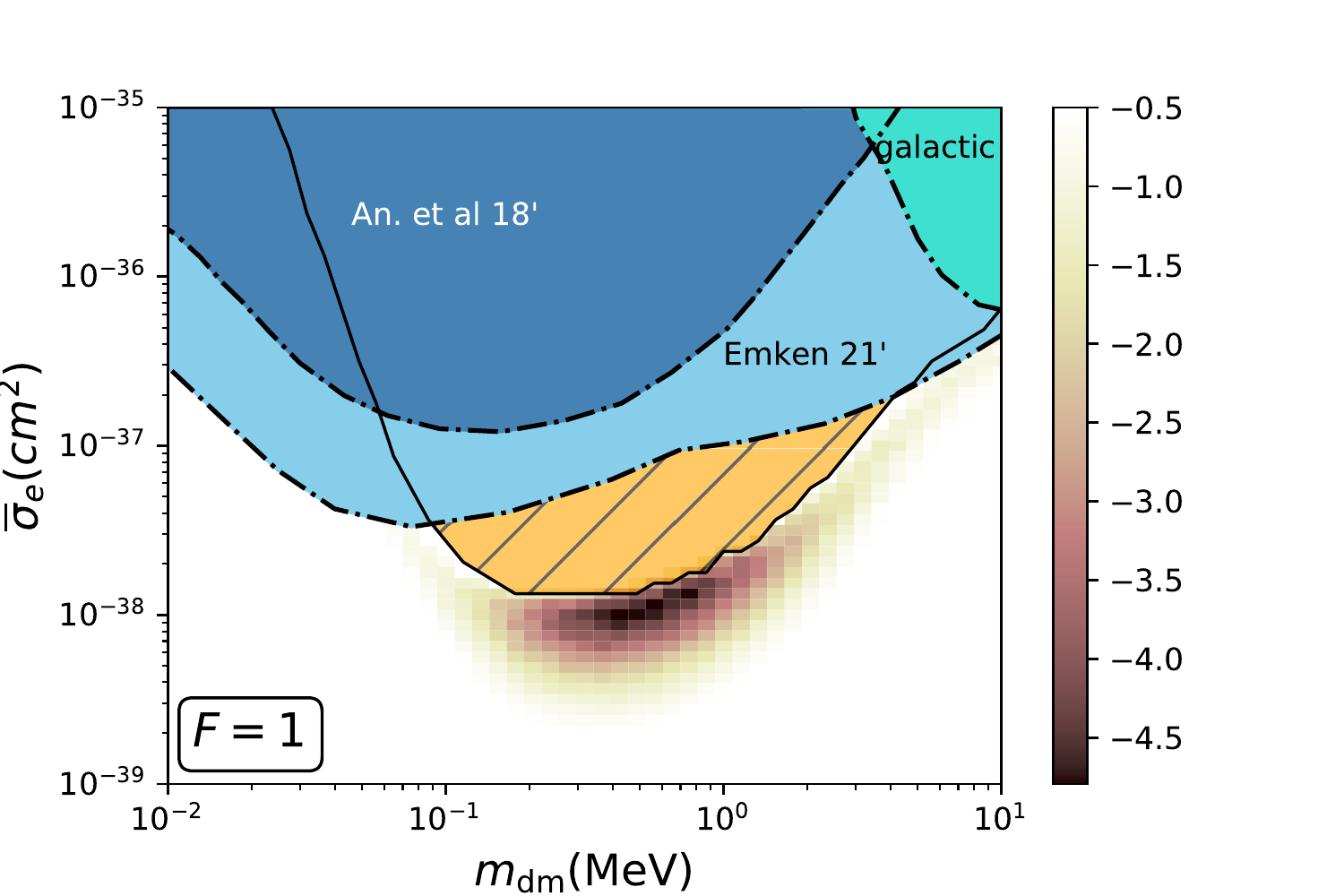} 
\includegraphics[width=0.45\textwidth]{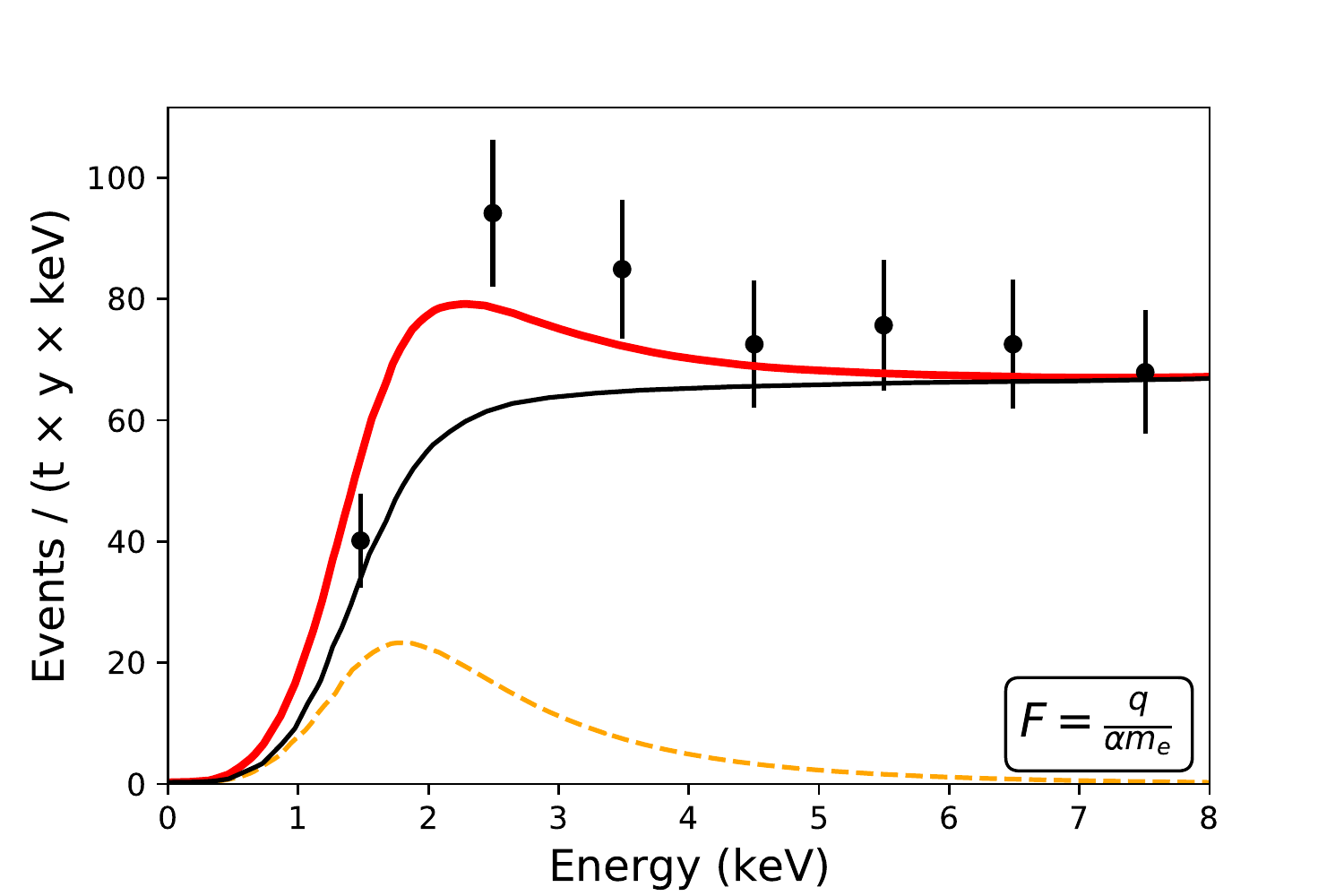} 
\includegraphics[width=0.45\textwidth]{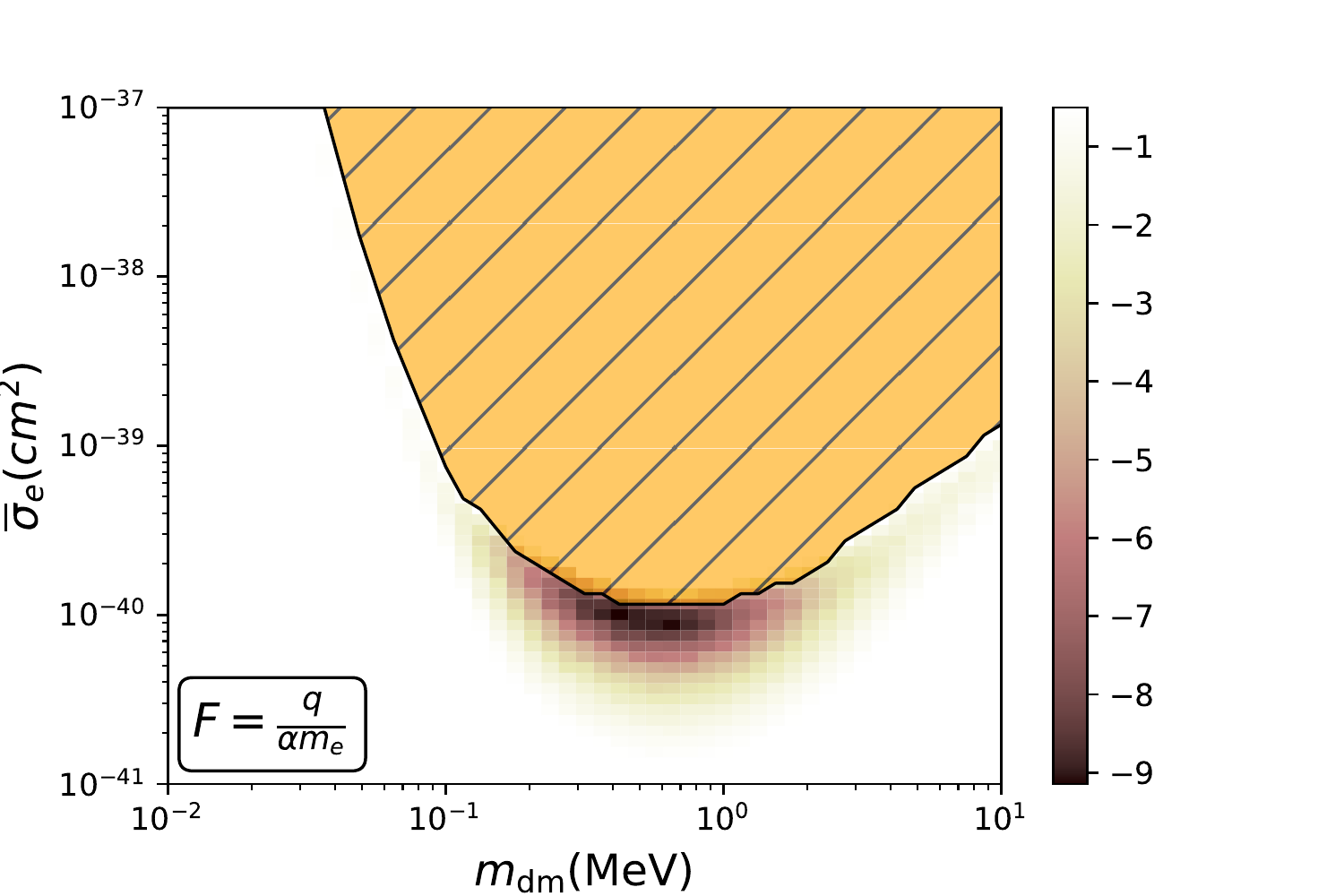} 
\includegraphics[width=0.45\textwidth]{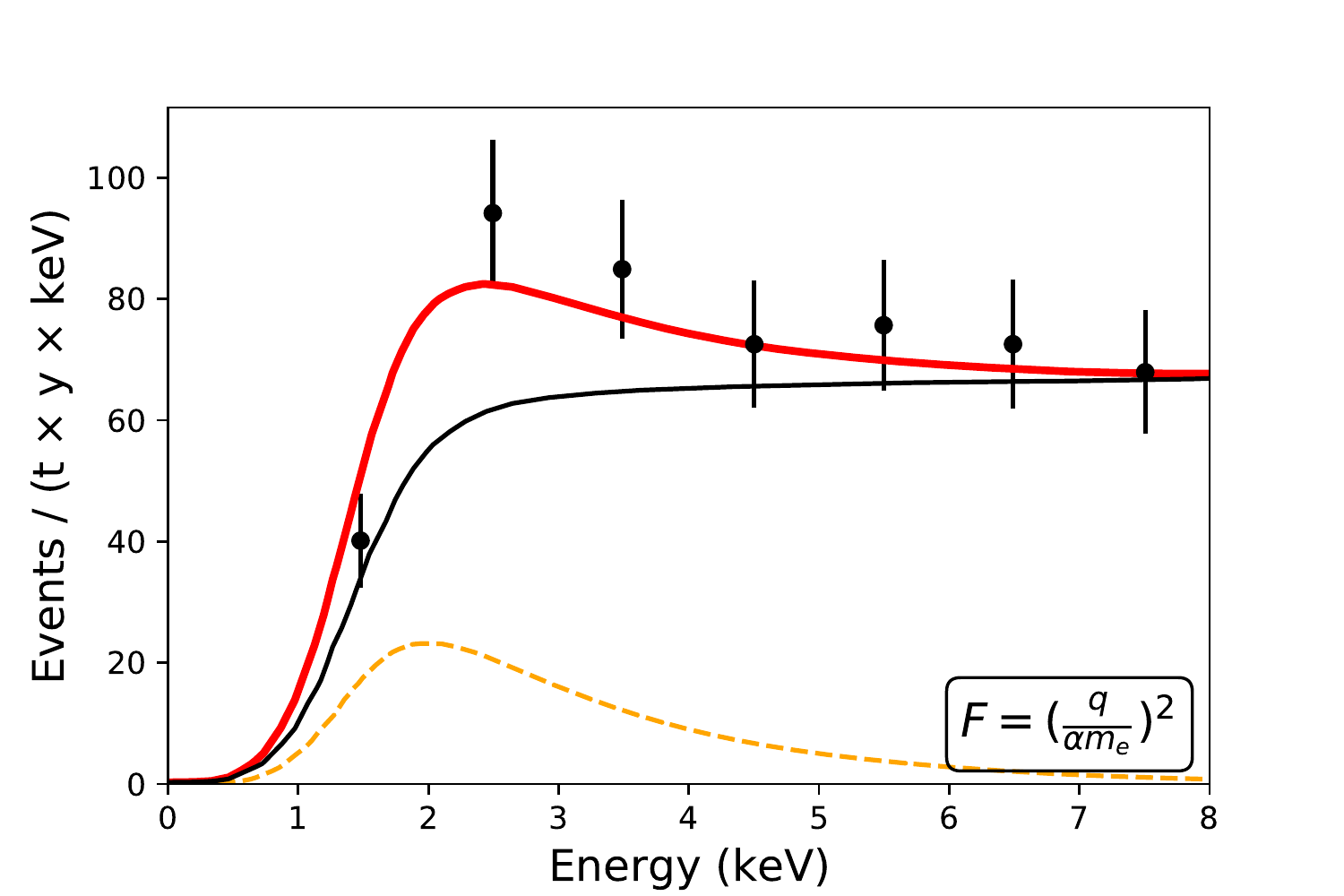} 
\includegraphics[width=0.45\textwidth]{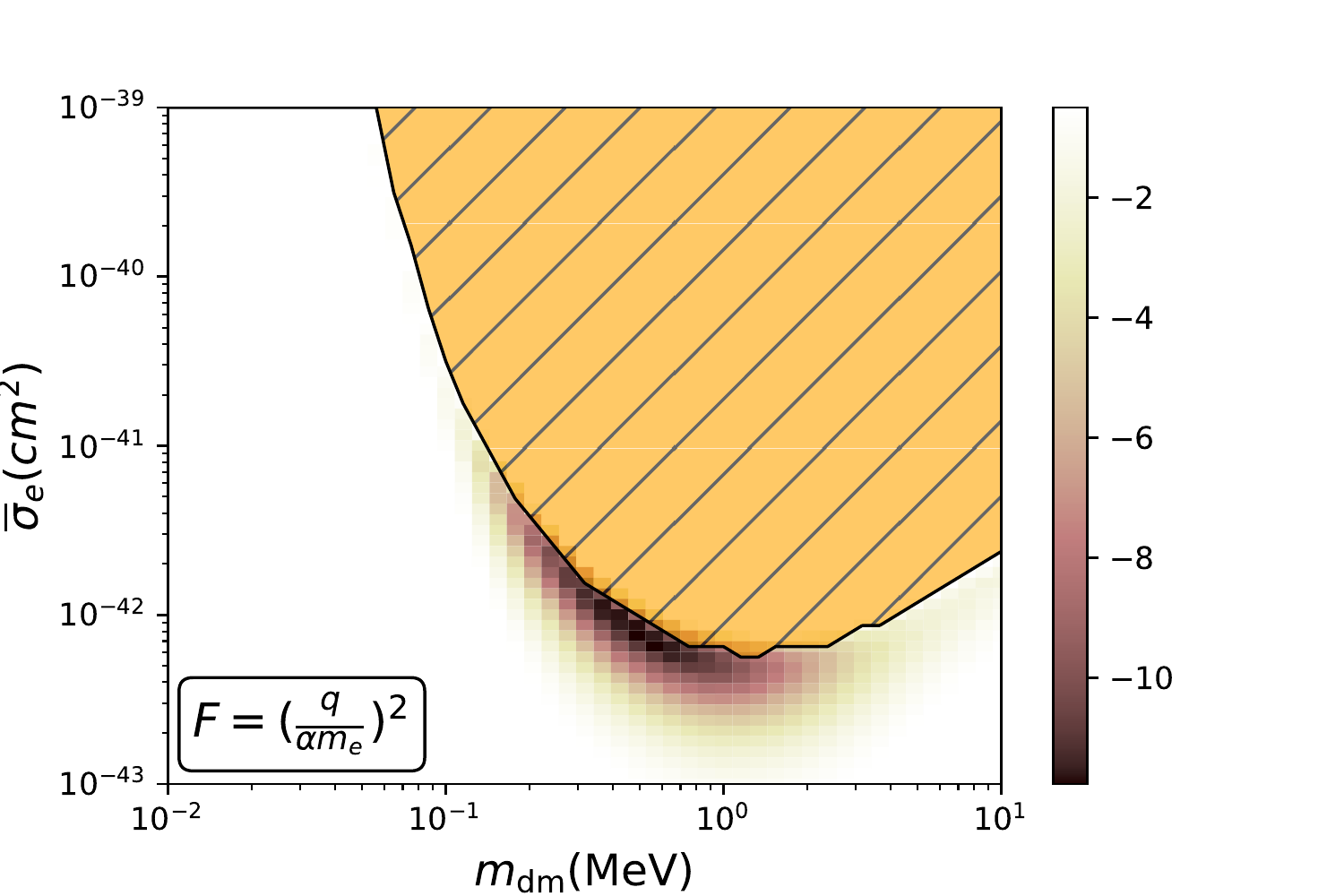} 
\caption{Left: Best-fit results to the XENON1T data when the Sun-heated 
DM-induced electron recoils are included, for $F(q)=1$ (top), 
$\frac{q}{\alpha m_\textrm{e}}$ (middle), and 
$(\frac{q}{\alpha m_\textrm{e}})^2$ (bottom), respectively. 
See the text for the best-fit model parameters. 
Right: 95\% exclusion regions (hatched orange areas) on the 
$m_{\rm dm}-\bar{\sigma}_\textrm{e}$ plane from the XENON1T 2020 data. 
For the $F(q)=1$ case, previous limits from Ref.~\cite{An:2017ojc} 
considering the Sun-heated DM scenario and the recent bounds from 
low-threshold SENSEI experiment \cite{Barak:2020fql} are also shown 
for comparison, as well as a bound based on the S2-only analysis
\cite{Aprile:2019xxb,Emken:2021lgc}.} 
\label{figboundandbfS}
\end{figure*}

We further convolve the event rate with a Gaussian energy resolution 
function with a width of $\sigma(E)=0.310\sqrt{\mathrm{keV}}\sqrt{E}+0.0037E$ 
and multiply the detection efficiency as given in Ref.~\cite{Aprile:2020tmw}. 
The best-fit results of the event rate distributions are shown in the left
panels Fig. \ref{figboundandbfS}, for the three form factors adopted here.
The best-fit model parameters and (logarithmic) likelihood ratios are
\begin{eqnarray}
m_{\rm dm}=0.42~\textrm{MeV},\ \bar{\sigma}_\textrm{e} = 1.0 \times 10^{-38} 
~\textrm{cm}^2,\ & 2 \ln \left(\mathcal{L}_{S+B} / \mathcal{L}_{B}\right) 
= 4.8, & F = 1; \nonumber\\
m_{\rm dm}=0.42~\textrm{MeV},\ \bar{\sigma}_\textrm{e}= 1.0 \times 10^{-40} 
~\textrm{cm}^2,\ & 2 \ln \left(\mathcal{L}_{S+B} / \mathcal{L}_{B}\right) 
= 9.3, & F = \frac{q}{\alpha m_\textrm{e}}; \nonumber\\
m_{\rm dm}=0.49~\textrm{MeV},\ \bar{\sigma}_\textrm{e}= 7.5 \times 10^{-43} 
~\textrm{cm}^2,\ & 2 \ln \left(\mathcal{L}_{S+B} / \mathcal{L}_{B}\right) 
= 12.1, & F = \left(\frac{q}{\alpha m_\textrm{e}}\right)^2.\label{bf}
\end{eqnarray}
In the above equation the likelihood function is defined as the Poisson 
likelihood $\mathcal{L} = \prod_{i=1}^7 e^{-\mu_i}\mu_i^{n_i} /n_i!$, where
$i$ denotes the $i$th energy bin, $n_i$ is the number of detected events, 
and $\mu_i$ is the expected number of events from the model. $S+B$ means the
signal$+$background model and $B$ means the background-only model. 
In this work the first 7 bins of the XENON1T data are considered. 
The logarithmic likelihood ratios of the fits give significance of the 
Sun-heated DM of $1.7\,\sigma$, $2.6\,\sigma$, and $3.0\,\sigma$, 
for $F(q)=1$, $\frac{q}{\alpha m_\textrm{e}}$, and 
$(\frac{q}{\alpha m_\textrm{e}})^2$, respectively, for 2 additional 
degrees of freedom.

As can be seen, for $F(q)=1$, the improvement of the fit compared with the
background hypothesis is not significant, due to the very soft energy 
spectrum of the Sun-heated DM. The results for 
$F=\frac{q}{\alpha m_\textrm{e}}$ and $F=(\frac{q}{\alpha m_\textrm{e}})^2$ 
are much better. This is expected, since the XENON1T data requires a 
relatively hard spectrum of the recoil events (the background expectation 
is consistent with the data in the first bin from 1 to 2 keV) and hence 
the DM spectrum. We further see that the best-fit values of the DM mass 
is indeed close to $m_e$, with a slight dependence on the form factor.

In the right panels of Fig. \ref{figboundandbfS} we show the distributions of
$-2\Delta\ln\mathcal{L}=-2\ln(\mathcal{L}_{S+B}/\mathcal{L}_{B})$ on the
$m_{\rm dm}-\bar{\sigma}_\textrm{e}$ plane. A negative value of $-2\Delta\ln\mathcal{L}$
means a favor of the model by the data. To be conservative, we also derive 
the $95\%$ upper bounds on the DM-electron cross section $\bar{\sigma}_\textrm{e}$, via requiring
$-2\Delta\ln\mathcal{L}=-2\ln(\mathcal{L}_{S+B}/\mathcal{L}_{S+B}^{\rm max})=2.71$,
for each fixed $m_{\rm dm}$. If a signal is indicated by the data, we start 
the scan of $\bar{\sigma}_\textrm{e}$ from the best-fit value to the larger side to derive 
the constraints. The $95\%$ exclusion limits are shown by the black dashed
lines in the right panels of Fig. \ref{figboundandbfS}. For the case of
$F(q)=1$, our results using the XENON1T 2020 data are more stringent that
that of Ref.~\cite{An:2017ojc} which considered the same Sun-heated DM model
but with XENON1T 2017 data. Compared with the results from the low-threshold
SENSEI experiment \cite{Barak:2020fql}, considering the DM scattering in the 
Sun can also effectively extend the sensitive range to lower DM masses.

For a given UV-completed model to realize the form factors, one 
requires the cutoff scale to be large enough to avoid the constraints 
from collider results \cite{Fox:2011fx}. 
For $F = \frac{q}{\alpha m_\textrm{e}}$, a dimension five operator 
$\phi^* \phi \bar{e} \gamma_5 e$ representing the interaction between 
scalar DM and electron spin, is discussed in \cite{Bloch:2020uzh} for 
$m_{\textrm{DM}} \simeq 100$ GeV where the cutoff scale can be hundreds 
of GeV that is consistent with both collider and electron Electric Dipole 
Moment (EDM) constraints. In our case, the lower dark matter mass can 
lead to an even larger cutoff around $10^6$ GeV \cite{Bloch:2020uzh}.
On the other hand, for $F = (\frac{q}{\alpha m_\textrm{e}})^2$, if one 
takes the dimension six operator $\bar{\chi}\gamma_{5}\chi\bar{e}\gamma_5 e$, 
the cutoff scale is always lower than $1$ GeV for the parameters in 
Eq. (\ref{bf}), which requires a more sophisticated model building 
to avoid the collider constraints. 

We note also that there are cosmological and astrophysical bounds 
on the DM-electron interactions, such as BBN, CMB, the overproduction 
of DM \cite{Lehmann:2020lcv}, and the supernova cooling 
\cite{DeRocco:2019jti, Chigusa:2020bgq}. These bounds are based on heavy 
mediator assumption with mass much higher than the typical energy scale 
around $10$~MeV. Discussion on DM coupled through lighter mediators with 
form factor $F=1$ were summarized in \cite{Knapen:2017xzo}, where the
HB stars exclude the mediator mass below $0.1$~MeV. Since the momentum 
exchange of the dominant scatterings considered in this work is relatively 
small ($\sim$ keV), we expect that there is a viable mass range of the 
mediator which is heavy enough that the contact interaction assumption 
in this work holds but is smaller than that required when the cosmological 
and supernova bounds apply. A late phase transition in the dark sector 
may modify the thermal history and relax the cosmological constraint 
as well (e.g., \cite{Chacko:2004cz,Davoudiasl:2017jke,Zhao:2017wmo}).

\section{Conclusion and discussion}

In this work we show that the electron recoil event excess detected by XENON1T 
\cite{Aprile:2020tmw} can be explained by MeV-scale DM particles interacting 
with electrons in the detector. While the slow (with velocity $\sim10^{-3}c$) 
DM particles in the Milky Way halo can not give electron recoils with energies 
of keV, the Sun plays a key role in heating a fraction of DM particles just to 
keV energies, leaving detectable signals in the detector. The goodness-of-fit
depends on the assumed form factor of the DM-electron scattering. 
For $F(q)\propto q^i$ ($i>0$), the model can fit the data reasonably well.
The best-fit DM mass is close to $m_e$, and the scattering cross section
$\bar{\sigma}_\textrm{e}$ is about $10^{-38}$ cm$^2$ (for $i=0$), 
$10^{-40}$ cm$^2$ (for $i=1$), and $10^{-42}$ cm$^2$ (for $i=2$). 
For $i=0$ and 1, these parameters can be consistent with other constraints 
(e.g., \cite{Ema:2018bih,Cappiello:2019qsw}).
However, for a strong $q$-dependence, other observations at higher energy 
scales may constrain the model. We emphasize that the physical process 
occurs in the Sun is the same as that in the detector, and thus no 
additional assumption is needed other than the DM-electron scattering. 
To be conservative, we also derive upper limits of the reference cross 
section $\bar{\sigma}_\textrm{e}$ between DM and electrons, for these 
different form factors. Note that for $i=1$ and 2, these constraints 
are presented for the first time. For $i=0$, our constraints 
are also stronger than previous works \cite{An:2017ojc,Barak:2020fql}.

Compared with other boosted DM models \cite{Kannike:2020agf,Fornal:2020npv}, 
the Sun-heated DM has a softer energy spectrum which would result in quite 
a few low-recoil-energy events. A rough estimate of the number of events 
due to the Sun-heated DM according to the best-fit model parameters in 
Eq. (\ref{bf}) gives consistent results with the XENON1T S2-only data 
\cite{Aprile:2019xxb}. Similar result was also shown recently in 
\cite{Emken:2021lgc}. Future direct detection experiments 
with lower thresholds or higher low-energy efficiencies would be able to 
distinguish this scenario from others. Furthermore, the direction sensitive 
direct detection experiments \cite{Mayet:2016zxu} may also directly test 
this model, with the Sun being the main source of such heated DM.

Cosmic rays in the Milky Way could also boost DM particles to high (or even 
very high) energies \cite{Cappiello:2018hsu,Bringmann:2018cvk,Yin:2018yjn}.
As already commented in Ref.~\cite{Kannike:2020agf}, the cosmic ray
electron boosted DM model seems to give conflicted results with that
of neutrino experiments, since the neutrino experiments are more
sensitive than the direct detection experiments for those electron
boosted DM \cite{Ema:2018bih}. For the scenario that DM particles 
are boosted by cosmic ray nuclei, which then interact with electrons
in the detector, the boosted DM fluxes seems to be also too low to
be consistent with the existing constraints. For example, taking
$F(q)=1$ and $\sigma_{\chi p}\sim10^{-31}$~cm$^2$ as an illustration, 
the peak flux of the boosted DM is about $10^{-6}$~cm$^{-2}$~s$^{-1}$
\cite{Bringmann:2018cvk,Ge:2020yuf}. For such a DM flux, the required 
cross section to account for the XENON1T excess events is 
${\mathcal O}(10^{-28})$ cm$^2$ \cite{Fornal:2020npv}, which exceeds 
significantly the current limits by neutrino experiments \cite{Ema:2018bih}.

\section*{Acknowledgements} 
We thank Shao-Feng Ge and Yue Zhao for helpful discussion.
Y.C. is supported by the China Postdoctoral Science Foundation under 
Grant No. 2020T130661, No. 2020M680688, the International Postdoctoral 
Exchange Fellowship Program, and by the National Natural Science 
Foundation of China (NSFC) under Grants No. 12047557.
J.S. is supported by the National Natural Science Foundation of China 
under Grants No. 12025507, No. 11690022, No.11947302; and is supported 
by the Strategic Priority Research Program and Key Research Program of 
Frontier Science of the Chinese Academy of Sciences under Grants 
No. XDB21010200, No. XDB23010000, and No. ZDBS-LY-7003.
Q.Y. is supported by the NSFC under Grants No.11722328, No.11851305,
Chinese Academy of Sciences, and the Program for Innovative Talents 
and Entrepreneur in Jiangsu. 
We acknowledge the use of HPC Cluster of ITP-CAS.

\bibliographystyle{unsrt}

\end{document}